\documentstyle[preprint,pra,floats,aps,psfig]{revtex}

\begin{document}

\title{Coherent control of nuclear spin isomers of\\ molecules:
The role of molecular motion\thanks{The work was presented at The
VIIth International Symposium on Magnetic Field and Spin Effects
in Chemistry and Related Phenomena, 15-20 July 2001, Japan}}
\author{P.L.~Chapovsky\thanks{E-mail: chapovsky@iae.nsk.su}}

\address{Institute of Automation and Electrometry,\\
          Russian Academy of Sciences, 630090 Novosibirsk, Russia}
         
\date{\today}

\draft
\maketitle

\begin{abstract}
Molecular center-of-mass motion is taken into account in the 
theory of coherent control of nuclear spin isomers of molecules. 
It is shown that infrared radiation resonant to the molecular 
rovibrational transition can substantially enrich nuclear spin 
isomers and speed up their conversion rate.
\end{abstract}

\vspace{2cm}

\section{Introduction}

Symmetrical molecules exist in the form of nuclear spin isomers
\cite{Landau81}. Most famous are the ortho and para isomers of H$_2$.
\cite{Farkas35}. On the other hand, spin isomers of molecules heavier 
than hydrogen are almost unknown. This is because their investigations 
were not possible for a long time due to the lack of enrichment methods. 
Now such methods start to appear that has advanced the field significantly 
(see the review \cite{Chap99ARPC}). Yet one needs more efficient 
enrichment methods in order to perform investigations 
of spin isomers of various molecules. It will also stimulate 
practical applications of spin isomers.  

It was proposed recently to use strong electromagnetic fields for
the spin isomer enrichment. In the first version of the method  
an enrichment is produced by the light-induced change of molecular
level populations \cite{Ilichov98CPL,Shalagin99JETPL}. In the second
method enrichment is due to combined action of the light-induced  
level populations change, level energy shift and
light-induced coherences \cite{Chap01PRA2,Chap01JPB}. 
The latter allows to refer to this method as coherent control of spin isomers.

Analysis of the isomer coherent control \cite{Chap01PRA2,Chap01JPB} was performed 
using simplified model of molecules interacting with a strong electromagnetic
field. Molecules were assumed to be at rest and have nondegenerate
states. The purpose of the present paper is to develop more general model 
which will account molecular center-of-mass motion. This can be essential 
because the process takes place in the gas phase at room temperature where 
light-molecular interaction is known to be affected by the molecular motion. 

\section{Qualitative picture}

Coherent control of spin isomers has been analysed for the case of  
microwave excitation of molecular rotational transitions \cite{Chap01PRA2} 
and for infrared excitation of
rovibrational transitions \cite{Chap01JPB}. Although each of these two
schemes has some advantages we will account molecular motion only for 
the case of rovibrational excitation because this scheme allows to 
achieve larger enrichment. Note also, that molecular motion is more
important for infrared excitation than for microwave excitation because
the latter has smaller Doppler broadening.

Let us start with the qualitative description of the phenomenon. We assume 
that the molecules have two nuclear spin states, ortho and para. 
This is the case, for example,
for the CH$_3$F molecules \cite{Landau81,Chap99ARPC}. Nuclear spin states are different
by the symmetry of nuclear spin wave function of equivalent nuclei in the molecule.
Long wavelength electromagnetic radiation cannot change the nuclear spin state
directly. An important point is that the spin conversion in polyatomic 
molecules is governed by intramolecular mixing of close (accidentally resonant)
ortho and para states (see Ref.~\cite{Chap99ARPC} for more details). 
Interaction of an external radiation with these particular states can affect
the spin conversion.

\begin{figure}[hb]
\centerline{\psfig
{figure=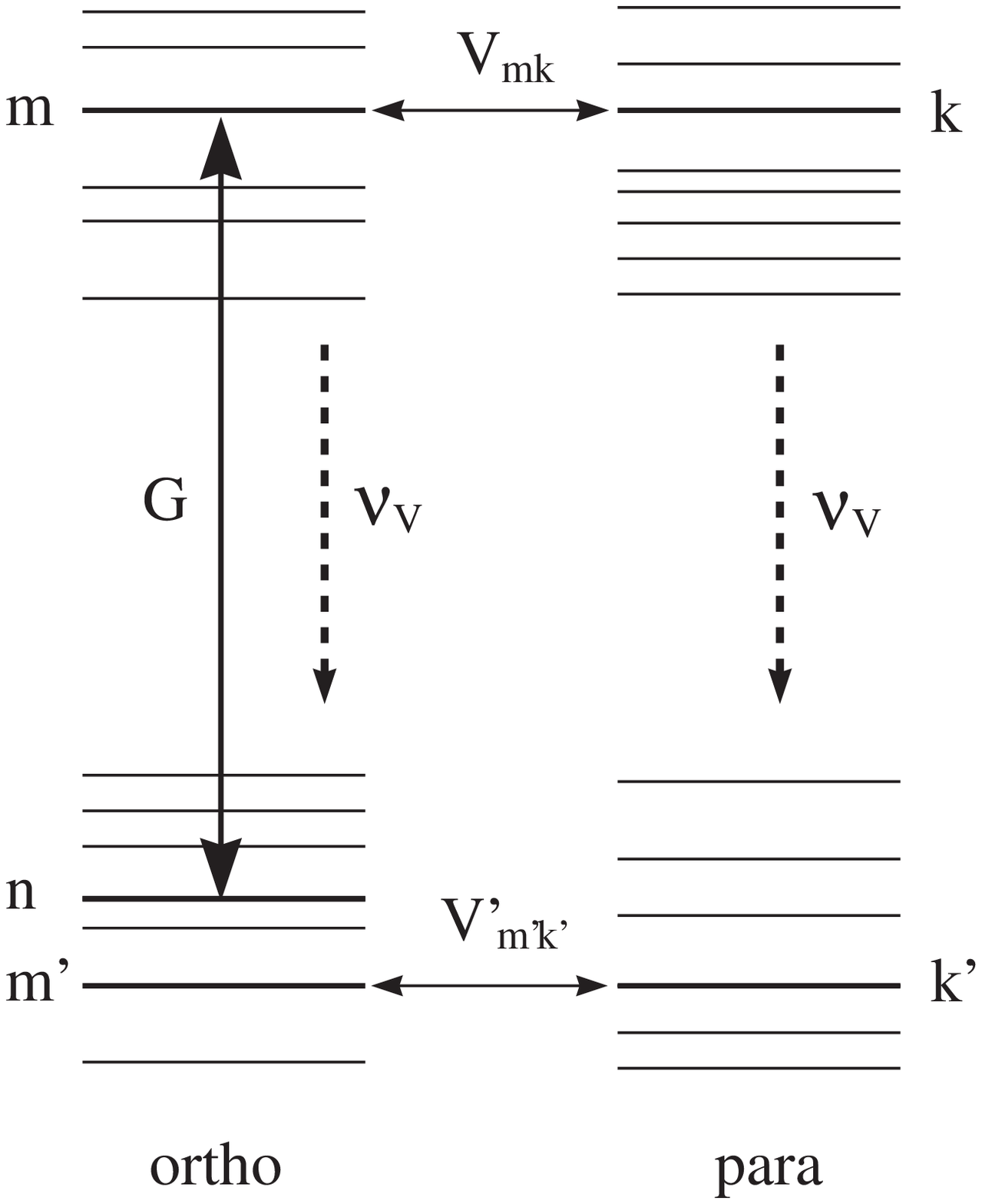,width=12cm}}
\vspace{0.5cm}
\caption{\sl Level scheme. Vertical solid line indicates infrared excitation.
Vibrational relaxation is shown by the dashed vertical lines. $V$ and $V'$ 
indicate the intramolecular ortho-para state mixing.}
\label{fig1}
\end{figure}

The level scheme for rovibrational excitation is shown 
in Fig.~\ref{fig1}. Similar to Ref.~\cite{Chap01JPB} we assume that each 
of the two vibrational states
(ground and excited) has one ortho-para level pair mixed by intramolecular
perturbation. The infrared radiation excites the rovibrational transition $m-n$ 
of the ortho molecules. By assumption, collisions cannot change spin state 
of the test molecules directly, i.e., collisional cross-section of
ortho-para transitions vanish. Suppose that the molecule is placed 
initially in the ortho subspace of the ground vibrational state.
Due to collisions the test molecule will undergo fast rotational
relaxation {\it inside} the ortho subspace. This will proceed until the molecule 
jumps to the state $m'$, which is mixed with the para state $k'$
by the intramolecular perturbation $\hat V'$ (see Fig.~\ref{fig1}), 
or to the state $n$ which is mixed with the para state $k$ by the 
combined action of the external field and intramolecular perturbation
$\hat V$. Admixture of a para state 
implies that the next collision can move the molecule to another para 
states and thus localize it inside the para subspace. The spin conversion
occurs. 

Vibrational relaxation in this enrichment scheme plays an important role.
After ortho-para transition in upper vibrational state the molecule relaxes
to the ground vibrational state. Consequently there is no back conversion through
the $m-k$ channel in excited vibrational state but back conversion  
occurs through the unperturbed $m'-k'$ channel. Steady state concentrations
of the spin isomers is determined by the ortho-to-para
conversion rate in the excited vibrational state and the para-to-ortho
conversion rate in the ground vibrational state. It is essential that conversion
through the $m-k$ channel can be substantially enhanced by electromagnetic
radiation.

It is clear that molecular center-of-mass motion affects the phenomenon 
because it makes the ortho-para mixing in upper vibrational state 
velocity selective. One obvious effect is that the
radiation does not interact with all molecules in the state $n$. 
Molecular motion influences also the level energy shift and coherences produced
in the molecule by external radiation. One needs quantitative account of
these effects in order to understand the real efficiency of the method.

\section{Rate equations}

Quantitative analysis of the problem can be performed with the help of 
kinetic equation for the density matrix, $\hat\rho$.
The molecular Hamiltonian consists of four terms,
\begin{equation}
     \hat H = \hat H_0 + \hbar\hat G + \hbar\hat V + \hbar\hat V'.
\label{H}
\end{equation} 
The main part, $\hat H_0$, has the eigen ortho and para states 
shown in Fig.~\ref{fig1}. $\hbar\hat G$ describes 
the molecular interactions with the external radiation that will be
taken in the form of monochromatic travelling wave,
\begin{equation}
\hat G = - ({\bf E}_0\hat{\bf d}/\hbar)\cos(\omega_Lt-\bf k\bf r),   
\label{g}
\end{equation}
where ${\bf E}_0$; $\omega_L$ and $\bf k$ are the amplitude, frequency 
and wave vector of the electromagnetic wave, respectively; $\hat {\bf d}$ 
is the operator of the molecular electric dipole moment. 
$\hat V$ and $\hat V'$ are the intramolecular perturbations
that mix the ortho and para states in excited and ground vibrational
states, respectively. These perturbations originate from the nuclear
spin-spin, or spin-rotation interactions 
\cite{Curl67JCP,Chap91PRA,Guskov95JETP,Ilisca98PRA}.

In the representation of the eigen states of $\hat H_0$ kinetic equation reads
\cite{Rautian79},
\begin{equation}
    \partial\text{\boldmath$\rho$}/\partial t +{\bf v}\cdot\nabla\text{\boldmath$\rho$} = 
    {\bf S} - i [{\bf G}+{\bf V}+{\bf V'},\text{\boldmath$\rho$}], 
\label{r1}
\end{equation}
where ${\bf S}$ is the collision integral; ${\bf v}$ is
the molecular center-of-mass velocity. We have considered molecular 
motion classically in this equation and neglected spontaneous
decay which is slow for molecular rovibrational transitions in comparison with 
collisional transitions. Account of molecular motion makes the matrices 
{\boldmath$\rho$} and ${\bf S}$  velocity dependent. The matrices
${\bf G}$, ${\bf V}$, and ${\bf V'}$ remain to be velocity independent.

Kinetic equation for the total concentration of molecules in one spin state,
e.g., ortho can be obtained from Eq.~(\ref{r1}), 
\begin{equation}
     \partial\rho_o/\partial t = 
     2Re\int i(\rho_{mk}V_{km}+\rho_{m'k'}V'_{k'm'})d{\bf v};\ \ \ 
     \rho_o = \sum_j\int\rho(j,{\bf v})d{\bf v};\ \ \ j\in {\text{ortho}}.
\label{ro}
\end{equation} 
Here $j$ indicates the rotational states of the molecule. These sates
are assumed to have no degeneracy. In Eq.~(\ref{ro})
a uniform spatial distribution of molecular density has been assumed. 
Collision integral did not enter into Eq.~(\ref{ro}) because by assumption 
collisions do not change the molecular spin state, i.e.,
$\sum_j\int S_{jj}d{\bf v}=0$, if $j\in$~ortho, or $j\in$~para. {\bf G} did
not enter into Eq.~(\ref{ro}) also because the matrix elements of {\bf G}   
off-diagonal in nuclear spin states vanish. 

The off-diagonal matrix elements $\rho_{mk}$ and $\rho_{m'k'}$ can be found
using the perturbation theory. We assume further the perturbations $\hat V$
and $\hat V'$ being small and consider zeroth- and first-order terms of the 
density matrix,
\begin{equation}
     \text{\boldmath$\rho$} = \text{\boldmath$\rho$}^{(0)} + 
                              \text{\boldmath$\rho$}^{(1)}.
\label{split}
\end{equation}

Collisions in our system will be described by the model standard
in the theory of light-molecular interaction. The off-diagonal elements 
of ${\bf S}$  have only decay terms,
\begin{equation}
     S_{jj'} = -\Gamma\rho_{jj'}; \ \ \  j\neq j'.
\label{S_off}
\end{equation}
The decoherence rates,
$\Gamma$, were taken equal for all off-diagonal elements of collision 
integral. Relaxation of the off-diagonal elements of the density matrix
is determined in molecular systems mainly by the level-population decay
which is thus assumed to be $j$-independent. Note that there 
are opposite examples, e.g., \cite{Trappeniers79JCP}.   

The diagonal elements of ${\bf S}$ can be expressed 
through the kernel of collision integral, $A$, as follows,
\begin{equation}
   S(\alpha) = \sum_{a_1j_1}\int A(\alpha|\alpha_1)\rho(\alpha_1)d{\bf v_1} -
   \rho(\alpha)\sum_{a_1j_1}\int A(\alpha_1|\alpha)d{\bf v_1};\ \ \ 
   \alpha\equiv \{a,j,{\bf v} \}.
\label{S}
\end{equation}
Here $a=e,g$ denotes the excited and ground vibrational states, respectively.
The nuclear spin index is omitted in Eq.~(\ref{S}) because there are 
no collisional ortho-para transitions and the kernels $A$ for ortho and para 
states are assumed to be the same. We will consider further the model of strong
collisions with the following collision kernel,
\begin{equation}
     A(\alpha|\alpha_1) = \nu_rw(j)\delta_{aa_1}\delta({\bf v}-{\bf v_1}) +
     \nu_tw(j)f({\bf v})\delta_{aa_1} + \nu_vw(j)f({\bf v})\delta_{a_1e}\delta_{ag}.
\label{A}
\end{equation}
Here $w(j)$ is the Boltzmann distribution of rotational state populations,
\begin{equation}
     w(j)=Z^{-1}\exp(-E_j/kT),
\label{B}
\end{equation}
where $Z$ is the rotational partition function for one vibrational state; 
$E_j$ is the rotational energy; $T$ is the
gas temperature; $k_B$ is the Boltzmann constant. We assume the same function 
$w(j)$ for the two vibrational states of both isomers.  $f({\bf v})$ is the
Maxwell distribution,
\begin{equation}
     f({\bf v}) =\frac{1}{\pi^{3/2}v_0^3}\exp(-\frac{{\bf v}^2}{v^2_0});\ \ \ 
     v_0=\sqrt{2k_BT/M}, 
\label{M}
\end{equation}
where $M$ is the molecular mass.

In Eq.~(\ref{A}) the three relaxation rates were introduced, rotational 
relaxation ($\nu_r$) without change of molecular velocity (velocity 
persistence effect) and vibrational state; translational relaxation 
($\nu_t$) that equilibrate also rotational
distribution but do not change the vibrational state;
vibrational relaxation ($\nu_v$) which is accompanied by 
velocity and rotational relaxations. The introduction of different relaxation 
rates for different degrees of freedom
makes the model of strong collisions more accurate and flexible. It allows
to adjust the model to the particular experimental conditions. 
Because of its simplicity model of strong collisions is often used in 
laser physics, see, e.g., \cite{Dykhne80JETP,Mironenko81IANS}.

\section{Zeroth-order theory}

For the zeroth-order term of the density matrix one has the following kinetic 
equation,
\begin{equation}
      \partial\text{\boldmath$\rho$}^{(0)}/\partial t +
      {\bf v}\cdot\nabla\text{\boldmath$\rho$}^{(0)} = 
      {\bf S}^{(0)}  - i [{\bf G},\text{\boldmath$\rho$}^{(0)}].
\label{ro0}
\end{equation}
Electromagnetic field excites only ortho molecules. Consequently, para
molecules remain at equilibrium in the zeroth-order perturbation theory. 
For the population of para molecules one has,
\begin{equation}
     \rho^{(0)}_p(g,j,{\bf v}) = 
     (N-\rho^{(0)}_o) w(j)f({\bf v});\ \ \ \rho^{(0)}_p(e,j,{\bf v}) = 0, 
\label{0rp}
\end{equation}
where $N$ is the total concentration of the test molecules; 
$\rho^{(0)}_o$ is the concentration of ortho molecules.

Zeroth-order theory for ortho molecules  
is in fact a familiar phenomenon of light-molecular interaction. 
There are numerous publications 
on this subject including many textbooks. Our zeroth-order theory 
is close to the approach \cite{Mironenko81IANS}, although the relaxation
scheme given by Eq.~(\ref{A}) is different from \cite{Mironenko81IANS}.
Eqs.~(\ref{S_off}),(\ref{S}),(\ref{A}), and (\ref{ro0}) allow to 
deduce the system of equations for the stationary populations of ortho molecules,
\begin{eqnarray}
\nu_e\rho^{(0)}_o(e,j,{\bf v}) & = & \nu_r w(j)\rho^{(0)}_o(e,{\bf v})
         + \nu_t w(j)f({\bf v})\rho^{(0)}_o(e)    
         + \rho^{(0)}_o p(\Omega,{\bf v})\,\delta_{jm}; \nonumber \\
\nu_g\rho^{(0)}_o(g,j,{\bf v}) & = &  \nu_r w(j)\rho^{(0)}_o(g,{\bf v})
          + \nu_t w(j)f({\bf v})\rho^{(0)}_o(g) +  \nonumber \\ 
          &&\nu_v w(j)f({\bf v})\rho^{(0)}_o(e) - 
          \rho^{(0)}_o p(\Omega,{\bf v})\,\delta_{jn}; \nonumber \\
 \rho^{(0)}_o p(\Omega,{\bf v}) & = & \frac{2\Gamma |G|^2}{\Gamma^2 + (\Omega-{\bf kv})^2}
                \left[\rho^{(0)}_o(g,n,{\bf v})  - \rho^{(0)}_o(e,m,{\bf v})\right],      
\label{sys0}
\end{eqnarray}
where $\nu_e=\nu_r+\nu_t+\nu_v$; $\nu_g=\nu_r+\nu_t$; 
$\rho^{(0)}_o(e)$ and $\rho^{(0)}_o(g)$ are the concentrations
of ortho molecules in excited and ground vibrational states;
$\rho^{(0)}_o(a,{\bf v})=\sum_j\rho^{(0)}_o(a,j,{\bf v});\ \ a=e,g$;
and $p(\Omega,{\bf v})$ is the probability of optical excitation. 
Eq.~(\ref{sys0}) is written in the rotating wave approximation.
Nonzero matrix element of ${\bf G}$ is given by,
\begin{equation}
     G_{mn} = - Ge^{i({\bf kr}-\Omega t)};\ \ 
     G\equiv E_0 \overline{d}_{mn}/2\hbar, 
\label{rwa}
\end{equation}
where $\Omega=\omega_L - \omega_{mn}$ is the radiation frequency
detuning from the absorption line center, $\omega_{mn}$; 
the bar over a symbol indicates a time-independent factor; 
$|G|$ is the Rabi frequency.

Solution of Eqs.~(\ref{sys0}) has no difficulty. Integration of 
Eqs.~(\ref{sys0}) over {\bf v} and summation over $j$ gives 
the system of algebraic equations for the distributions
$\rho^{(0)}_o(a,j,{\bf v})$, $\rho^{(0)}_o(a,{\bf v})$, and $\rho^{(0)}_o(a)$
that can be easily solved. The level populations
can be expressed as the sum of three terms, equilibrium in $j$ and ${\bf v}$,
equilibrium in $j$, and nonequilibrium in $j$ and ${\bf v}$,
\begin{eqnarray}
 \rho^{(0)}_o(e,j,{\bf v})& = & \rho^{(0)}_o\{p(\Omega)\tau_{ev}w(j)f({\bf v})+
      p(\Omega,{\bf v})[\tau_{et}w(j)+\tau_e\delta_{jm}]\} \nonumber \\
 \rho^{(0)}_o(g,j,{\bf v})& = & \rho^{(0)}_o\{w(j)f({\bf v})[1-
      p(\Omega)\tau_{gv}]-p(\Omega,{\bf v})[\tau_{gt}w(j)+\tau_g\delta_{jn}]\}.      
\label{sol0}
\end{eqnarray}
here $p(\Omega)=\int p(\Omega,{\bf v})d{\bf v}$ and,
\begin{equation}
 p(\Omega,{\bf v})  =  \frac{w(n)\Gamma^2_Bf({\bf v})}
    {\tau_1[\Gamma^2_B+(\Omega-{\bf kv})^2][1+1/\kappa+Y(\Omega)\tau_2/\tau_1]}.       
\label{pov}
\end{equation}
In this equation, $\Gamma_B=\Gamma\sqrt{1+\kappa}$ is the power dependent 
homogeneous linewidth (Bennett width) and $\kappa=2|G|^2\tau_1\Gamma^{-1}$ 
is the saturation parameter. The relaxation times are expressed through
the collisional parameters of the kernel $A$ from Eq.~(\ref{A}),
\begin{eqnarray}
 \tau_{ev} & = & \nu^{-1}_v-(\nu_v+\nu_t)^{-1};\ \ \ 
         \tau_{et}=(\nu_v+\nu_t)^{-1}-\nu_e^{-1};\ \ \ \tau_e=\nu_e^{-1} \nonumber \\
 \tau_{gv} & = & \nu^{-1}_v-\nu^{-1}_t;\ \ \ \tau_{gt}=\nu^{-1}_t -\nu^{-1}_g ;\ \ \ 
         \tau_g=\nu^{-1}_g;\ \ \ \nonumber \\
    \tau_1 & = & \tau_{et}w(m)+\tau_{gt}w(n)+\tau_e+\tau_g;\ \ \ 
         \tau_2=\tau_{ev}w(m)+\tau_{gv}w(n).     
\label{tau}
\end{eqnarray}
The function,
\begin{equation}
     Y(\Omega) = \sqrt{\pi}yReW(x+iy);\ \ \ x=\Omega/kv_0;\ \ \ y=\Gamma_B/kv_0,
\label{Y}
\end{equation}
is the Foigt profile of the absorption coefficient. $W(z)$ is the probability
integral,
\begin{equation}
     W(z) = e^{-z^2}\left[1+\frac{2i}{\sqrt{\pi}}\int^z_0e^{u^2}du\right].
\label{W}
\end{equation}

Similar to the calculation of the diagonal matrix elements of 
{\boldmath$\rho$}$^{(0)}_o$, one can find the off-diagonal density matrix 
element, $\rho^{(0)}_o(n|m;{\bf v})$ which amplitude is equal to,
\begin{equation}
     \overline\rho^{(0)}_o(n|m;{\bf v}) = -i\rho^{(0)}_o
     \frac{p(\Omega,{\bf v})}{2G\Gamma}[\Gamma-i(\Omega-{\bf kv})].
\label{ro_off}
\end{equation} 

\section{First order theory}

In zeroth-order perturbation theory, one neglects perturbations
$\hat V$ and $\hat V'$. It implies that there are no coherences
between ortho and para states, $\rho^{(0)}_{mk}=0$;  
$\rho^{(0)}_{m'k'}=0$. Consequently, one has,
\begin{equation}
     \partial\rho_o/\partial t = 2Re \int 
     i(\rho^{(1)}_{mk}V_{km}+\rho^{(1)}_{m'k'}V'_{k'm'})d{\bf v},
\label{ro2}
\end{equation} 
instead of Eq.~(\ref{ro}). Thus the spin conversion appears
in the second order approximation.

The first order correction to the 
density matrix, {\boldmath$\rho$}$^{(1)}$, is determined by the equation, 
\begin{equation}
     \partial\text{\boldmath$\rho$}^{(1)}/\partial t + 
     {\bf v}\cdot\nabla\text{\boldmath$\rho$}^{(1)} = {\bf S}^{(1)} 
        - i [{\bf G},\text{\boldmath$\rho$}^{(1)}] 
        - i [{\bf V} + {\bf V'},\text{\boldmath$\rho$}^{(0)}].
\label{ro1}
\end{equation}
For $\rho^{(1)}_{m'k'}$ one has from this equation,
\begin{equation}
     \rho^{(1)}_{m'k'} = \frac{-iV'_{m'k'}}{\Gamma + i\omega'}
     [\rho^{(0)}_p(g,k',{\bf v})-\rho^{(0)}_o(g,m',{\bf v})],  
\label{1rho}
\end{equation}
where $\omega'\equiv\omega_{m'k'}$. The density matrix element
$\rho^{(1)}_{mk}$ can be obtained from equations
which are deduced from Eq.~(\ref{ro1}),
\begin{eqnarray}
\partial\rho^{(1)}_{mk}/\partial t + {\bf v}\cdot\nabla\rho^{(1)}_{mk} + 
          \Gamma\rho^{(1)}_{mk} + iG_{mn}\rho^{(1)}_{nk}
          & = &  iV_{mk} \rho^{(0)}_o(e,m,{\bf v}); \nonumber \\
\partial\rho^{(1)}_{nk}/\partial t  + {\bf v}\cdot\nabla\rho^{(1)}_{nk} + 
          \Gamma\rho^{(1)}_{nk} + iG_{nm}\rho^{(1)}_{mk} 
          & = & iV_{mk} \rho^{(0)}_o(n|m;{\bf v}).    
\label{sys1}
\end{eqnarray}
Substitutions, 
     $V_{mk}=\overline{V}e^{i\omega t},\ \ (\omega\equiv\omega_{mk});\ \ 
     \rho^{(1)}_{mk} = \overline{\rho}^{(1)}_{mk}e^{i\omega t};\ \ 
     \rho^{(1)}_{nk} = \overline{\rho}^{(1)}_{nk}
     e^{i[(\Omega-\omega)t-{\bf kv}]}$,
transform Eqs.~(\ref{sys1}) to algebraic equations from which one 
finds $\rho^{(1)}_{mk}$. Using this result for $\rho^{(1)}_{mk}$ and 
$\rho^{(1)}_{m'k'}$ 
from Eq.~(\ref{1rho}) one has the rate equation (\ref{ro2}) in the form,
\begin{eqnarray}
 \frac{\partial\rho_o}{\partial t} & = & \frac{2\Gamma|V'|^2}{\Gamma^2+\omega'^2}
    \left[\rho^{(0)}_p(g,k')-\rho^{(0)}_o(g,m') \right] - \nonumber \\ 
  &&2|V|^2 Re\int\frac{[\Gamma+i(\Omega+\omega-{\bf kv})]\rho^{(0)}_o(e,m,{\bf v}) +
    iG\overline\rho^{(0)}_o(n|m;{\bf v})}{(\Gamma+i\omega)
    [\Gamma+i(\Omega+\omega-{\bf kv})]+|G|^2}d{\bf v}.    
\label{ro3}
\end{eqnarray}
This rate equation has the same general structure as in 
\cite{Chap01JPB}. On the other hand the contributions from the level population in
upper state, $\rho^{(0)}_o(e,m,{\bf v})$, and from the light-induced coherence,
$\rho^{(0)}_o(n|m;{\bf v})$, are more complicated now due to their
velocity dependencies.  

\section{Enrichment and conversion}

Similar to Ref.~\cite{Chap01JPB}, the rate equation (\ref{ro3}) is  
presented in the form, 
\begin{equation}
     \partial\rho_o/\partial t  =  N\gamma'_{op} - \rho_o\gamma;\ \ \ 
     \gamma  \equiv  \gamma'_{op} + \gamma'_{po} - \gamma'_n + \gamma_n + \gamma_{coh}.
\label{final}
\end{equation}
We have neglected in the right-hand side of Eq.~(\ref{final}) the small
difference between $\rho^{(0)}_o$ and the total concentration of
ortho molecules $\rho_o$. In Eq.~(\ref{final}) the following partial conversion
rates were introduced. The field independent rates,
\begin{equation}
     \gamma'_{op}=\frac{2\Gamma|V'|^2}{\Gamma^2+\omega'^2}w(k');\ \ \ 
     \gamma'_{po}=\frac{2\Gamma|V'|^2}{\Gamma^2+\omega'^2}w(m').
\label{g'}
\end{equation}
$\gamma_{free}\equiv \gamma'_{op} + \gamma'_{po}$, is
the isomer conversion rate without an external field. The rates $\gamma'_{op}$ and 
$\gamma'_{po}$ coincide with the analogous rates in \cite{Chap01JPB}.
This is not surprising because intramolecular ortho-para state mixing is
velocity independent. The field dependent contribution to the
conversion rate,
\begin{equation}
     \gamma'_n = \gamma'_{po}p(\Omega)w(m')(\tau_{gv}+\tau_{gt}),
\label{g'n}
\end{equation}
is also similar to that in \cite{Chap01JPB} except for the new definition of 
relaxation parameters. This rate enters to the total rate $\gamma$ in 
Eq.~(\ref{final}) with minus sign because it is caused by the light-induced
population depletion of the ground vibrational state of ortho molecules.
The last two contributions to $\gamma$ are more complicated now.
The ``noncoherent'' term appears due to the level population, 
$\rho^{(0)}_o(e,m,{\bf v})$, in Eq.~(\ref{ro3}),
\begin{equation}
     \gamma_n = 2|V|^2 Re\int\frac{[\Gamma+i(\Omega+\omega-{\bf kv})]
     [p(\Omega)\tau_{ev}w(m)f({\bf v})+p(\Omega,{\bf v})(\tau_{et}w(m)+\tau_e)]}
     {(\Gamma+i\omega)[\Gamma+i(\Omega+\omega-{\bf kv})]+|G|^2}d{\bf v}.
\label{gn}
\end{equation}
The ``coherent'' term, $\gamma_{coh}$, originates from
$\overline\rho^{(0)}_o(n|m;{\bf v})$, in Eq.~(\ref{ro3}),
\begin{equation}
     \gamma_{coh} = |V|^2\Gamma^{-1}Re\int
     \frac{p(\Omega,{\bf v})[\Gamma-i(\Omega-{\bf kv})]}
     {(\Gamma+i\omega)[\Gamma+i(\Omega+\omega-{\bf kv})]+|G|^2}d{\bf v}.
\label{gcoh}
\end{equation}

General expressions for the conversion rates are rather difficult to
analyse. It will be done elsewhere. Below we compare the results 
\cite{Chap01JPB} with the present calculations. Solution to 
Eq.~(\ref{final}) can be presented as,
\begin{equation}
    \rho_o = \overline\rho_o + (\rho_o(0)-\overline\rho_o)\exp(-\gamma t);\ \ \  
    \overline\rho_o = N\gamma'_{op}/\gamma.
\label{overro}
\end{equation}
Here $\overline\rho_o$ is the stationary concentration of ortho molecules.
Without an external radiation (at the instant $t=0$), the equilibrium
concentration of para molecules reads,
\begin{equation}
     \rho_p(0)=N-\rho_o(0)=N\gamma'_{po}/\gamma_{free}= N/2,
\label{rop}
\end{equation}   
if the Boltzmann factors are assumed to be equal, $w(k')=w(m')$. 
This implies $\gamma'_{op}=\gamma'_{po}$ (see Eq.~(\ref{g'})). 
A stationary enrichment of para molecules is defined as,
\begin{equation}
     \beta \equiv \frac{\overline\rho_p}{\rho_p(0)}-1 = 
     1-2\frac{\gamma'_{op}}{\gamma}.
\label{bp}
\end{equation}
One can see from this equation that the external field produces 
an enrichment of para isomers if $\gamma \neq \gamma_{free}$.

We assume in further analysis the following parameters,
the Doppler width of the absorbing transition, $kv_0=30$~MHz;
$\omega=100$~MHz, $\omega'=130$~MHz, $V'_{m'k'}=5$~kHz, 
$\Gamma/P=2\cdot10^8$~s$^{-1}$/Torr and the 
Boltzmann factors of the states $m'$, $k'$, $m$, and $k$ all equal
$10^{-2}$. This set of parameters gives the field free conversion rate, 
$\gamma_{free}=10^{-2}$~s$^{-1}$/Torr, which coincides with the  
conversion rate in $^{13}$CH$_3$F. Nuclear spin conversion in these
molecules is governed by quantum relaxation (see the review \cite{Chap99ARPC}). 
The translational, rotational and vibrational relaxation rates will be taken equal, 
$\nu_t=0.1\Gamma$, $\nu_r=0.1\Gamma$, and $\nu_v=0.01\Gamma$, respectively.

\begin{figure}[htb]
\centerline{\psfig
{figure=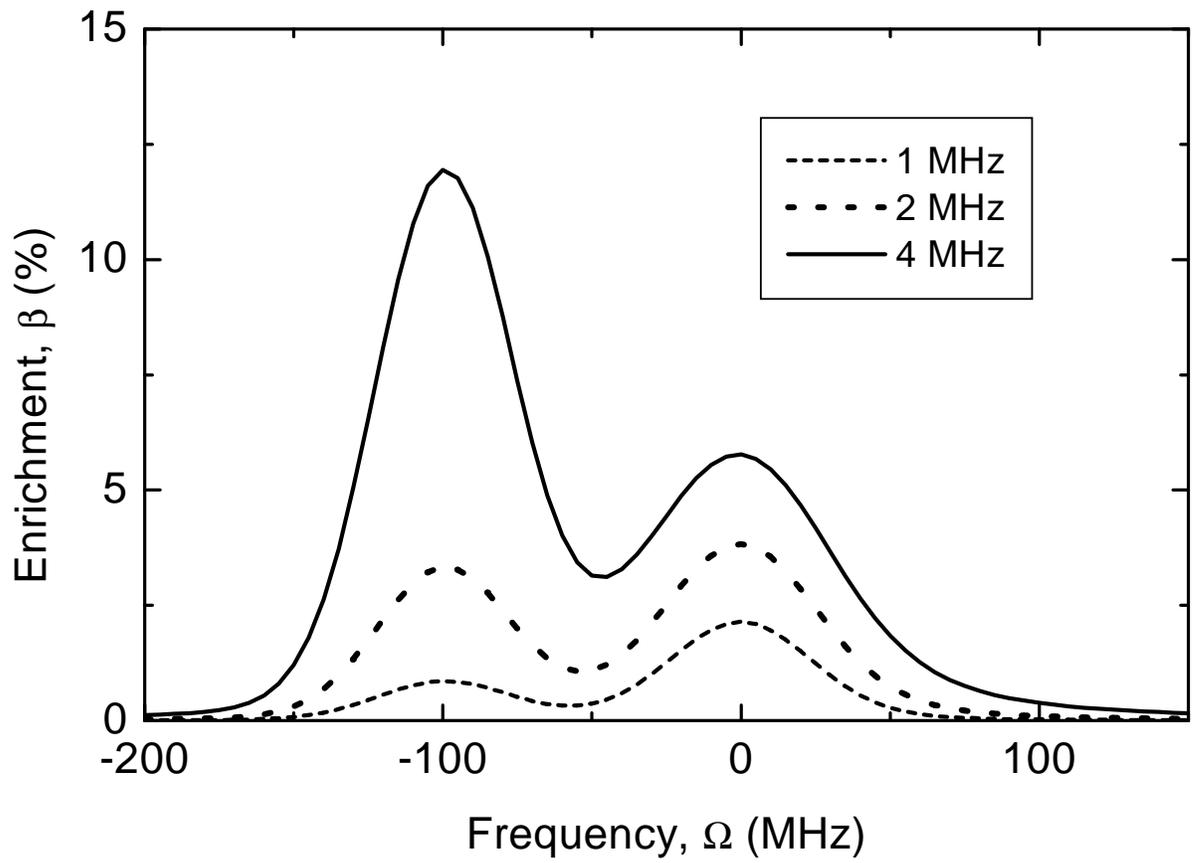,height=14cm}}
\vspace{2cm}
\caption{\sl Frequency dependence of the enrichment of para molecules, 
$\beta$, at the Rabi frequencies $|G|=1$, 2, and 4~MHz.}
\label{fig2}
\end{figure}

Close similarity between noncoherent, $\gamma_n$, and coherent, $\gamma_{coh}$,
terms (see Eqs.~(\ref{gn}) and (\ref{gcoh})) suggests that their
contributions  to the total rate, $\gamma$, are rather difficult
to separate. It is indeed so for a wide range of parameters. On the
other hand, at some conditions it is possible. For example, it can
be done for weak external fields, thus small $G$. In this case one has,
similar to \cite{Chap01JPB}, two peaks in enrichment at frequencies 
$\Omega\simeq-\omega$ and $\Omega=0$, see Fig.~\ref{fig2}. An obvious 
difference is that these two peaks are much wider now having the 
Doppler width, $kv_0$. The data shown in this figure
correspond to $V_{mk}=3$~kHz, and $\Gamma=2$~MHz. The peak at $\Omega=0$
appears because the excitation probability, $p(\Omega)$, has maximum at
$\Omega=0$. This peak has positive contribution from the rate $\gamma_n$
and negative contributions from the rates $\gamma'_n$ and $\gamma_{coh}$. 
At high $G$ the main contribution originated from $\gamma_n$ reaches the value, 
\[
\frac{\gamma_n}{\gamma_{free}}\sim \frac{1}{2}\left(\frac{V\omega'}
{V'\omega}\right)^2\frac{\tau_{ev}w(m)+\tau_{et}w(m)+\tau_e}{\tau_1+\tau_2},
\]
which constitutes $\simeq 15\%$ in our numerical example. 
The homogeneous width of the peak at $\Omega=0$ is equal to $\Gamma_B$. 
Consequently, this peak has rather large power broadening.  
The peak at $\Omega=0$ in isomer enrichment was predicted in
\cite{Ilichov98CPL} by considering only the level population effects. 

The ``off-resonant'' peak at $\Omega\simeq -\omega$ appears 
because the external field splits the ortho state $m$ into 
two components and one sublevel crosses  the para state $k$ at 
this frequency of the external field (see Ref.~\cite{Chap01JPB} 
for more details). At low $G$ this peak is determined by
$\gamma_{coh}$. When $G$ increases the peak at $\Omega\simeq -\omega$ 
grows to much bigger values than the peak at $\Omega=0$. 
At high $G$ the main contribution to the $\Omega\simeq -\omega$ 
peak comes from the rate $\gamma_n$. Position of this peak is given by 
$-\omega[1-|G|^2/(\Gamma^2+\omega^2)]$, 
and the homogeneous width is equal to,
$\Gamma[1+|G|^2/(\Gamma^2+\omega^2)]$,
thus this peak has much smaller homogeneous width than the peak
at $\Omega=0$ if $\omega\gg\Gamma$.

\begin{figure}[htb]
\centerline{\psfig
{figure=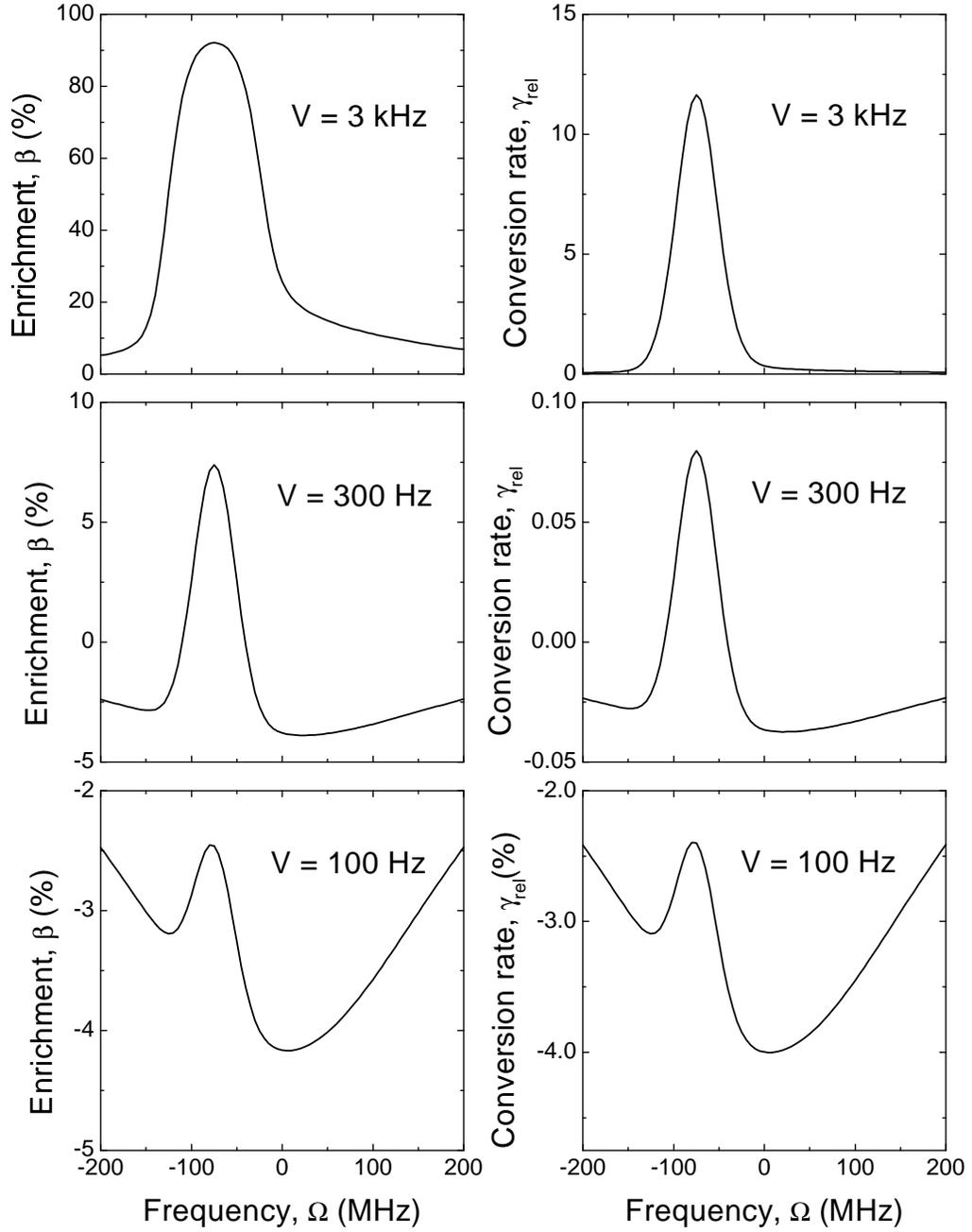,height=20cm}}
\vspace{2cm}
\caption{\sl Enrichment of para molecules, $\beta$, and conversion rate, 
$\gamma_{rel}$, for the three values of $V_{mk}$. The wide negative structures
in the middle and low panels (the width equal $\Gamma_B$) are mainly due to
$\gamma'_n$.}
\label{fig3}
\end{figure}

The data shown in Fig.~\ref{fig3} correspond to strong optical field, 
$G=50$~MHz, and three values of $V_{mk}$. $\Gamma$ was taken equal 2~MHz. 
One can see, that strong optical field is able to convert
almost all molecules to the para state if $V_{mk}\simeq V_{m'k'}$.
Thus relatively weak (3~kHz) coupling in upper state produces 
macroscopic effect, viz., almost complete enrichment of spin isomers.
In the case of much weaker coupling
in upper state, enrichment is still significant. For example, if the
perturbation in upper state, $V_{mk}=100$~Hz, one has the 
enrichment, $\beta\simeq1$\%. Enrichment at this level can 
easily be measured.  The enrichment peak at 
$\Omega\simeq -\omega$ is narrow (having almost the Doppler width) 
and thus can be distinguished from much wider structures
(the width $\simeq \Gamma_B$) caused by the $\Omega=0$ peaks.

Conversion rate in the system is equal to $\gamma$, see Eq.~(\ref{overro}). 
It is convenient to characterize it in relative units,
\begin{equation}
     \gamma_{rel} = \gamma/\gamma_{free} - 1.
\label{grel}
\end{equation}
Thus, $\gamma_{rel}=0$ without an external field.
At low $G$, conversion rate, like enrichment, has two peaks in its frequency
dependence. If Rabi frequency, $|G|$, is large and the ortho-para couplings
in upper and low states have the same order of magnitude, conversion can be
significantly enhanced (Fig.~\ref{fig3}, upper panel), 
although this enhancement is smaller than in the case of the molecules at rest
\cite{Chap01JPB}. This 
enhancement appears because of the crossing of ortho and para states in
upper vibrational state by external field. The main difference between the
present case and Ref.~\cite{Chap01JPB} is that the peaks in conversion rate
are much wider now having the Doppler width, $kv_0$. 

\section{Conclusions}

We have developed the model of coherent control of nuclear
spin isomers taking into account the molecular center-of-mass motion. This
model is more realistic than the model \cite{Chap01JPB} in which the molecules
were assumed to be at rest. It was shown that the main conclusions of \cite{Chap01JPB}
still hold. One has two peaks in the enrichment and conversion rate dependences
on the laser radiation frequency. One can achieve a substantial enrichment and 
speed up the conversion rate. It is also valid that this
enrichment scheme can be used to access very weak perturbations in
molecules. On the other hand, molecular motion causes some decrease in sensitivity
of the method. For example, 1\% enrichment can be achieved for the
intramolecular perturbation equal $V\simeq100$~Hz in the present model, 
in comparison with 1\% enrichment at $V\simeq30$~Hz in Ref.~\cite{Chap01JPB}.
This is not a significant difference. But even this loss in sensitivity can 
probably be avoided in special excitation scheme, e.g., by using a multifrequency 
excitation instead of a monochromatic excitation. This needs further investigations.

Search of proper resonances will be one of the main difficulty in 
practical implementation of the proposed enrichment scheme. An existence 
of the off-resonance peak in the isomer enrichment has particular 
importance. First, one does not need to search for an exact 
coincidence of the ortho and para states in upper
vibrational state. The energy gap can be compensated by detuning
the laser frequency. Second, the excitation in the off-resonant peak
does produce even higher enrichment than in the resonant peak but does not
produce significant population in upper vibrational state. The latter
helps not to populate the excited vibrational state of other spin isomer and
consequently suppress the spurious back conversion through the upper channel.  

\section*{Acknowledgments}

This work was made possible by financial support from the  
Russian Foundation for Basic Research (RFBR), grant No.~01-03-32905
and from The Institute of Physical and Chemical Research (RIKEN), 
Japan.

\newpage

\newpage

\newpage

\end{document}